\documentclass [11pt]{article}
\usepackage{epsf} 
\usepackage{amsmath}
\usepackage{amssymb}
\usepackage{epsfig}
\usepackage{latexsym}
\usepackage{amsfonts}
\usepackage{varioref}
\usepackage{ifthen}
\usepackage{graphicx}
\usepackage{amssymb,amsmath,amsthm}
\begin{document}
\parindent 0mm 
\setlength{\parskip}{\baselineskip} 
\thispagestyle{empty}
\pagenumbering{arabic} 
\setcounter{page}{0}
\mbox{ }
\rightline{UCT-TP-296/13}
\newline
\rightline{July 2013}
\newline
\vspace{0.2cm}
\begin{center}
{\Large {\bf Bottonium in QCD at finite temperature}}
\end{center}
\vspace{.1cm}
\begin{center}
{\bf C. A. Dominguez}$^{(a)}$,  {\bf M. Loewe}$^{(a),(b)}$, 
 {\bf Y. Zhang}$^{(a)}$
\end{center}
\begin{center}
$^{(a)}$Centre for Theoretical \& Mathematical Physics,University of
Cape Town, Rondebosch 7700, South Africa

$^{(b)}$Facultad de F\'{i}sica, Pontificia Universidad Cat\'{o}lica de Chile, Casilla 306, Santiago 22, Chile

\end{center}
\vspace{0.3cm}
\begin{center}
\textbf{Abstract}
\end{center}
Vector ($\Upsilon$) and pseudoscalar ($\eta_b$) bottonium ground states are studied at finite temperature in the framework of thermal Hilbert moment QCD sum rules. The mass, the onset of perturbative QCD in the complex squared energy plane, $s_0$, the leptonic decay constant, and the total width are determined as a function of the temperature. Results in both channels show very little temperature dependence of the mass and of $s_0$, in line with expectations. However, the width and the leptonic decay constant exhibit a very strong $T$-dependence. The former increases with increasing temperature, as in the case of light- and heavy-light-quark systems, but close to the critical temperature, $T_c$, and for $T/T_c \simeq 0.9$ it drops dramatically approaching its value at $T=0$, as obtained recently in this framework for charmonium states.  The leptonic decay constant is basically a monotonically increasing function of the temperature, also as obtained in the charmonium channel . These results are interpreted as the survival of these bottonium states above $T_c$, in line with lattice QCD results.\\ 
\newpage
\noindent
\section{INTRODUCTION}
\noindent
The abundant literature on the extension of the method of QCD sum rules (QCDSR) \cite{REV} to finite temperature \cite{QCDSRT} leads to the following scenario. In the complex squared energy s-plane,  Cauchy's theorem allows to relate hadronic parameters, e.g. masses, couplings and widths, to QCD parameters such as vacuum condensates and $s_0$, the onset of perturbative QCD (PQCD). For light-quark, and heavy-light-quark systems $s_0(T)$ and the current coupling $f(T)$ have been found to be monotonically decreasing functions of $T$, with the width $\Gamma(T)$ increasing substantially with increasing temperature, and the mass showing a small increase or decrease, depending on the channel (for recent results see \cite{recent} and references therein). 
This behaviour is consistent with quark-gluon deconfinement and chiral symmetry restoration at a critical temperature $T_c \simeq 200 \,{\mbox{MeV}}$. In fact, as $s_0(T)$ approaches the hadronic threshold, poles and resonances begin to disappear from the spectrum as the coupling decreases and the width increases. The thermal behaviour of the mass is irrelevant in this scenario, as it only provides information on the real part of the Green function. An intriguing exception has been the results in the charmonium channel \cite{JPSI}-\cite{ETAC}, where the vector ($J/\psi$), scalar ($\chi_c$), and pseudoscalar ($\eta_c$) ground states appear to survive beyond $T_c$. Indeed, the width of these states  initially increases with temperature, but it it reverses this trend close to $T_c$ where it decreases dramatically approaching its value at $T=0$. The coupling is basically a monotonically increasing function of $T$. These results are in qualitative agreement with lattice QCD (LQCD) \cite{LQCD}. A quantitative, point by point comparison is not feasible as the definition of the critical temperature, and of the deconfinement parameter in LQCD is different from that in QCDSR. Nevertheless, keeping this difference in mind, it is rewarding to find such a qualitative agreement.\\

In this paper we extend the QCDSR analysis of charmonium \cite{JPSI}-\cite{ETAC} to ground state  bottonium in the vector ($\Upsilon$) and the pseudoscalar ($\eta_b$) channels. This is particularly relevant in view of LQCD results for the temperature dependence of the $\Upsilon$ and the $\eta_b$ width \cite{BOTT}. It would probably be the first time that a T-dependent parameter determined from QCDSR can be directly compared  with that from LQCD. Our results for the thermal mass and PQCD threshold in both channels show a very slight decrease with increasing temperature. Such a behaviour was already obtained for charmonium  \cite{JPSI}-\cite{ETAC}, and it is due to $s_0(0)$  being very close to the hadronic/QCD threshold. The $T$-dependence of the width and coupling of $\Upsilon$ and $\eta_b$ also resemble that of the  $J/\psi$ and $\eta_c$, respectively, thus suggesting the survival of bottonium beyond $T_c$.  In fact, we find that the thermal behaviour of  $\Gamma(T)$ as a function of $T/T_c$, in both bottonium channels, is in qualitative agreement with LQCD results as after increasing at first it then drops dramatically near $T \simeq T_c$. It should be recalled that in the QCDSR approach once $s_0(T)$ reaches the hadronic/QCD threshold there are no longer solutions to the sum rules, as there is no support for the hadronic/QCD integrals which vanish identically.  It must also be kept in mind that in applications of  QCDSR  for heavy-heavy-quark systems $T_c$ is basically determined by the vanishing of the gluon condensate. Given these differences with LQCD it is reassuring to find agreement on the qualitative temperature behaviour of the width.
\newpage

\section{QCD SUM RULES}
In order to study vector and  pseudoscalar bottonium we consider the thermal current correlator

\begin{equation}
\Pi (q^{2},T)   = i \, \int\; d^{4}  x \; e^{i q x} \; \;\theta(x_0)\;
<<|[ J(x) \;, \; J{\dagger}(0)]|>> \;,
\end{equation}

where $J(x) = : \bar{Q}(x) \Gamma Q(x):$, with $\Gamma =\gamma_\mu$ ($\Gamma = \gamma_5$) for the vector (pseudoscalar) channel,  and $Q(x)$ is the heavy quark field. The vacuum to vacuum matrix element above is the Gibbs average

\begin{equation}
<< A \cdot B>> = \sum_n exp(-E_n/T) <n| A \cdot B|n> / Tr (exp(-H/T)) \;,
\end{equation}

where $|n>$ is any complete set of eigenstates of the (QCD) Hamiltonian, H. We  adopt the quark-gluon basis, as this allows for a straightforward and smooth extension of the QCD sum rule method to finite temperature \cite{QCDSRT}. In the case of heavy-quark systems it has been customary to use Hilbert moment QCD sum rules \cite{REV}, e.g. in the vector channel (requiring a once subtracted dispersion relation)

\begin{equation}
\varphi_N(Q_0^2,T) \equiv \frac{1}{N!}\, \Bigl(- \frac{d}{dQ^2}\Bigr)^N \Pi(Q^2,T)|_{Q^2=Q_0^2} = \frac{1}{\pi}
\int_{0}^{\infty} \frac{ds}{(s+Q_0^2)^{(N+1)}}\,  Im \,\Pi(s,T)\; , 
\end{equation}

where $N = 1,2,...$, and $Q_0^2 \geq 0$ is an external four-momentum squared to be considered as a free parameter \cite{RRYNP}. Using Cauchy's theorem in the complex squared energy s-plane, leading to quark-hadron duality, the Hilbert moments become Finite Energy QCD sum rules (FESR) \cite{REV}, i.e.

\begin{equation}
\varphi_N(Q_0^2, T)|_{HAD} =  \varphi_N(Q_0^2,T)|_{QCD} \;,
\end{equation}

where the hadronic and the QCD moments are 

\begin{equation}
\varphi_N(Q_0^2,T)|_{HAD} \equiv \frac{1}{\pi}
\int_{0}^{s_0(T)}\frac{ds}{(s+Q_0^2)^{(N+1)}}\, Im \,\Pi(s,T)|_{HAD} \;,
\end{equation}

\begin{equation}
 \varphi_N(Q_0^2,T)|_{QCD} \equiv \frac{1}{\pi}
\int_{4 m_b^2}^{s_0(T)}\frac{ds}{(s+Q_0^2)^{(N+1)}}\, Im \,\Pi_{PQCD}(s,T) 
+ \varphi_N(Q_0^2,T)|_{NP}  \;,
\end{equation}

with $m_b$ the bottom-quark mass, $Im \Pi_{PQCD}(s,T)$ the PQCD spectral function, and $\varphi_N(Q_0^2,T)|_{NP}$  the non perturbative moments involving vacuum condensates in the operator product expansion (OPE) of the current correlator. For heavy-heavy quark Green functions the gluon condensate is the leading term in this expansion. In the sequel, the quark mass is considered independent of the temperature, a good approximation \cite{mbT} for $T < 200 - 250 \;{\mbox{MeV}}$.

Starting with the vector channel, the hadronic spectral function is parametrized as usual in terms of the  ground state resonance, i.e. $\Upsilon(1S)$, followed by a continuum given by PQCD starting at a threshold $s_0$, the radius of the integration contour in the complex s-plane. At finite temperature this ansatz is a much better approximation than at $T=0$ because $s_0(T)$ is expected to decrease monotonically with increasing temperature. The ground state must be considered in finite width, $\Gamma = \Gamma(T)$, as this is a crucial parameter providing information on deconfinement. The hadronic Hilbert moments then become

\begin{equation}
	\varphi_N|_V(Q_0^2,T)|_{HAD} = \frac{2}{\pi}f_V^2(T) M_V(T) \Gamma_V(T) \int_{0}^{s_0(T)} \frac{ds}{(s+Q_0^2)^{N+1}}\frac{1}{[s-M_V^2(T)]^2 + M_V^2(T)\Gamma_V^2(T)},
\end{equation}

where the leptonic decay constant is defined as

\begin{equation}
<0| V_\mu(0) | V(k)> = \sqrt{2}\; M_V \;f_V \;\epsilon_\mu \;.
\end{equation}

At finite temperature there is in principle an additional hadronic contribution \cite{JPSI}-\cite{ETAC} arising from a cut centered at the origin in the complex energy $\omega$-plane, of length $-|{\bf q}| \leq \omega \leq +|{\bf q}|$, with space-like $q^2 = \omega^2 - {\bf q}^2 < 0$. This is interpreted as arising from the vector current scattering off  heavy-light quark pseudoscalar mesons (B-mesons). It has been shown in \cite{JPSI} for the case of charmonium that this term is exponentially suppressed. Given the mass gap between charmonium and bottonium, this term is absolutely negligible here. Turning to the QCD sector, the PQCD moments in the time-like (annihilation) region, $\varphi_N^a(Q_0^2,T)|_{PQCD}$, are \cite{JPSI}

\begin{equation}
	\varphi_N^a|_V(Q_0^2,T)|_{PQCD} = \frac{1}{8\pi^2}\int_{4m_b^2}^{s_0(T)}\frac{ds}{(s + Q_0^2)^{N+1}}\; v(s)\; [3-v(s)^2]\;\left[1-2n_F\left(\left|\frac{\sqrt{s}}{2 T}\right|\right)\right]\;,
\end{equation}

where $ v^2(s) = 1 - 4 m_b^2/s$, $s = \omega^2 - \mathbf{q}^2 \geq 4 m_b^2$, and  $n_F(z) = (1+e^z)^{-1}$ is the Fermi thermal function. In the space-like region there is a non-negligible contribution from the center cut in the complex energy plane, $\varphi_N^s(Q_0^2,T)|_{PQCD}$, given by \cite{JPSI}

\begin{equation}
	\varphi_N^s|_V(Q_0^2,T)|_{PQCD} = \frac{2}{\pi^2}\frac{1}{(Q_0^2)^{N+1}}\left[m_b^2 \;n_F(m_b/T) + 2\int_{m_b}^\infty  y \; n_F(y/T)\; dy\right].
\end{equation}

As in all applications of QCD sum rules at finite temperature, we consider all QCD correlation functions only to leading order in PQCD (one-loop approximation) . In fact,  QCD sum rules are valid in the whole range $0 \leq T \leq T_c$, a region where  thermal PQCD beyond one-loop order is not valid. In fact, the strong coupling, $\alpha_s(Q^2,T)$, involves two scales, $\Lambda_{QCD}$ as well as $T_c$, so that PQCD is expected to be valid for $Q^2 >> \Lambda_{QCD}^2$, as well as for $T > T_c$. This two-scale problem was identified long ago \cite{QCDSRT}, but it remains unsolved. From a practical point of view this has  very limited impact on thermal   QCD sum rule applications, as results are not intended to be of high precision. In addition, by determining the ratio of parameters at finite and at zero $T$ as a function of $T/T_c$ one effectively minimizes the uncertainty. 

\begin{figure}[ht]
	\centerline{
	\mbox{\includegraphics[scale=0.75]{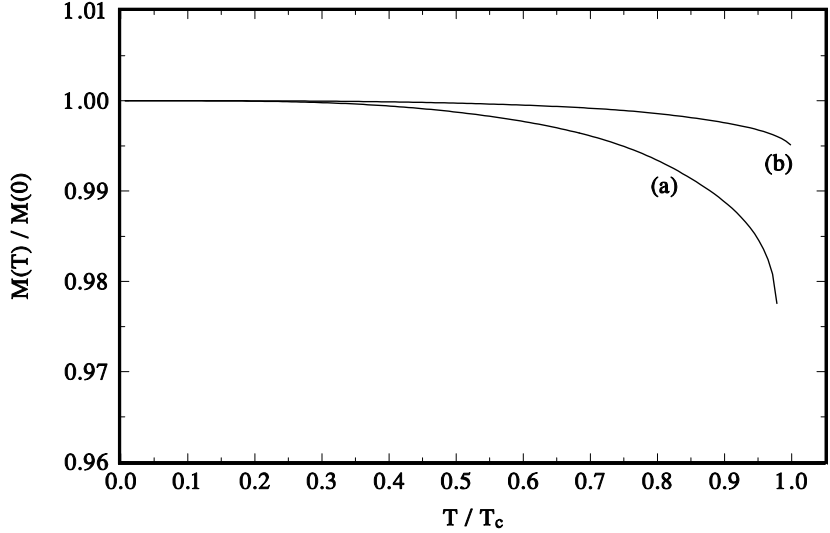}}
	}
	\caption{The ratios of hadron masses, $M(T)/M(0)$, against $T/T_c$ in the vector channel ($\Upsilon$), curve (a), and in the pseudoscalar channel ($\eta_b$), curve (b).}
	\label{M}
\end{figure}

The Hilbert moments of the leading non perturbative correction in the OPE, i.e the gluon condensate, are given by \cite{RRYNP}

\begin{eqnarray}
	\varphi_N|_V(Q_0^2,T)|_{NP} &=& -\frac{1}{3} \; \frac{2^N N(N+1)^2(N+2)(N+3)(N-1)!}{(2N+5)(2N+3)!!}\frac{1}{\left[4m_b^2(1+\xi)\right]^{N+2}}
\nonumber \\ [.3cm] &\times&	F\left(N+2, -\frac{1}{2}, N + \frac{7}{2},\rho\right)\left\langle\frac{\alpha_s}{\pi}G^2\right\rangle_T \;,
\end{eqnarray}

where

\begin{equation}
	F(a,b,c;z) = \sum_{N=0}^{\infty} \frac{(a)_N (b)_N} {(c)_N} \frac{z^N}{N!} 
\end{equation}

is the hypergeometric function with  $(a)_N = a(a+1)(a+2)...(a+N-1)$, 
$\xi \equiv \frac{Q_0^2}{4m_b^2}$, $\rho \equiv \frac{\xi}{1+\xi}$, and $\langle\frac{\alpha_s}{\pi}G^2\rangle_T$ is
the thermal gluon condensate, i.e. the dimension $d=4$ leading term in the OPE. At finite temperature there are, in principle, additional contributions to the OPE arising from nondiagonal (Lorentz noninvariant) condensates. Both gluonic and nongluonic terms can be safely ignored, as discussed in detail in \cite{JPSI}. \\

\begin{figure}[ht]
	\centerline{
	\mbox{\includegraphics[scale=0.75]{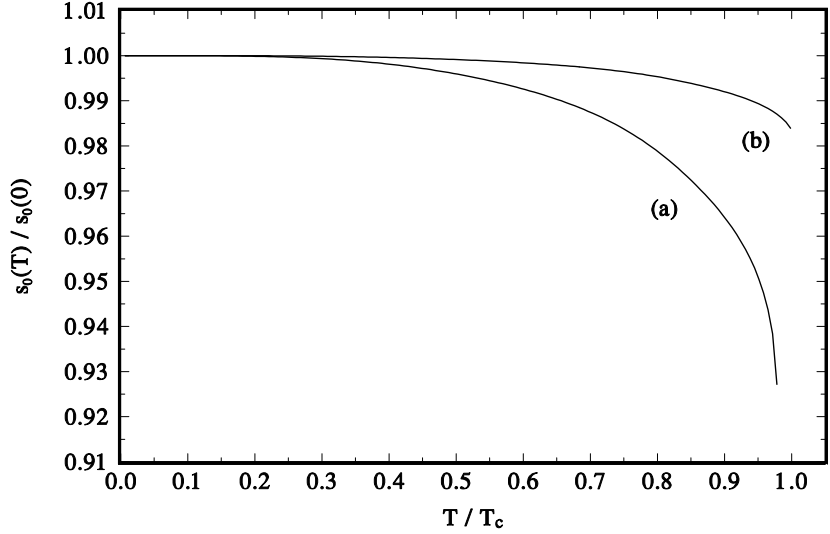}}
	}
	\caption{The ratios $s_{0}(T)/s_{0}(0)$ against $T/T_c$ in the vector channel ($\Upsilon$), curve (a), and in the pseudoscalar channel ($\eta_b$), curve (b).}
	\label{S0}
\end{figure}

Turning to the pseudoscalar channel, the hadronic Hilbert moments, requiring now a twice subtracted dispersion relation, are 

\begin{equation}
	\varphi_N|_P(Q_0^2,T)|_{HAD}= \frac{2}{\pi}f_P^2(T) M_P^3(T) \Gamma_P(T) \int_{0}^{s_0(T)} \frac{ds}{(s +Q_0^2)^{N+2}}\frac{1}{[s-M_P^2(T)]^2 + M_P^2(T) \Gamma_P^2(T)} \;,
\end{equation}

where the leptonic decay constant is defined as

\begin{equation}
\langle 0| J_5(0)|0 \rangle = \sqrt{2} \,f_P \, M_P^2 \,.
\end{equation}

The PQCD moments corresponding to the time-like (annihilation) region become

\begin{equation}
	\varphi_N^a|_P(Q_0^2,T)|_{PQCD} = \frac{3}{8\pi^2}\int_{4m_b^2}^{s_0(T)}  \frac{ds}{(s + Q_0^2)^{N+2}} \, s\, v(s) \left[1-2n_F\left(\frac{\sqrt{s}}{2T}\right)\right].
\end{equation}

\begin{figure}[ht]
	\centerline{
	\mbox{\includegraphics[scale=0.75]{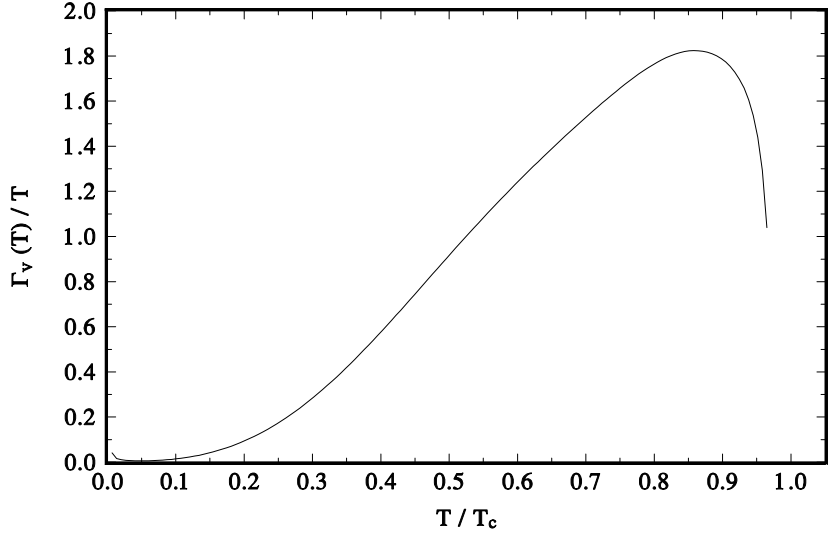}}
	}
	\caption{The ratio of width to temperature, $\Gamma_V(T)/T$, against $T/T_c$ in the vector channel ($\Upsilon$).} 
	\label{GV2}
\end{figure}

The PQCD contribution in the space-like (scattering) region vanishes identically, as shown in \cite{ETAC}. The corresponding hadronic term is loop suppressed, as it would involve a two-loop diagram instead of a tree-level one. The Hilbert moments of the gluon condensate are now

\begin{eqnarray}
	&&\varphi_{N}|_P(Q_0^2,T)|_{NP} = -\frac{3}{8\pi^2}\frac{2^{(N+1)}N!}{(4m_b^2)^{(N+1)}}\frac{1}{(1+\xi)^{N+2}} \frac{(N+1)(N+2)(N+3)(N+4)}{(2N+5)(2N+3)!!} \nonumber\\ 
	&\times& \left[F\left(N+2,-\frac{3}{2},N+\frac{7}{2};\rho\right) - \frac{6}{N+4}F\left(N+2,-\frac{1}{2},N+\frac{7}{2};\rho\right)\right]\Phi(T),
\end{eqnarray}
where 

\begin{equation}
 \Phi(T) \equiv \frac{4\pi^2}{9} \frac{1}{(4m_b^2)^2}
 \langle \frac{\alpha_s}{\pi} G^2 \rangle (T) \;.
\end{equation} 

\begin{figure}[ht]
	\centerline{
	\mbox{\includegraphics[scale=0.75]{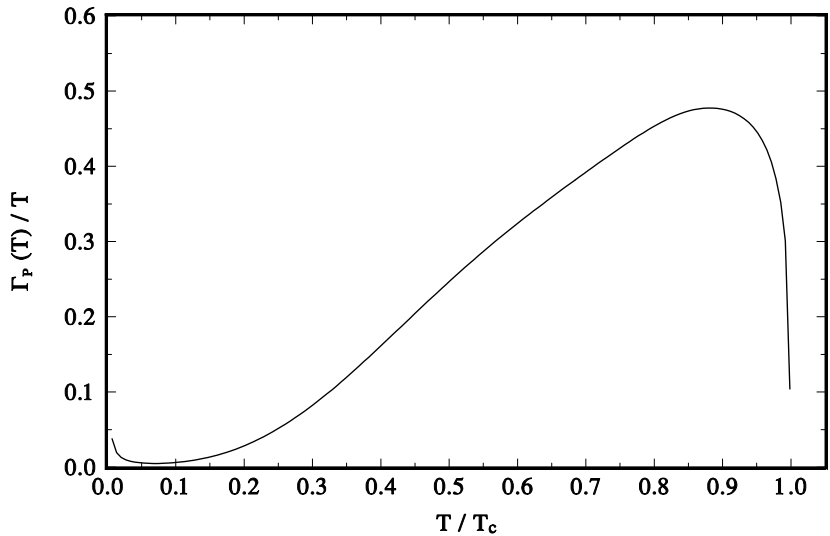}}
	}
	\caption{The ratio $\Gamma_P(T)/T$ against $T/T_c$ in the pseudoscalar channel ($\eta_b$).}
	\label{GPS}
\end{figure}

\begin{figure}[ht]
	\centerline{
	\mbox{\includegraphics[scale=0.70]{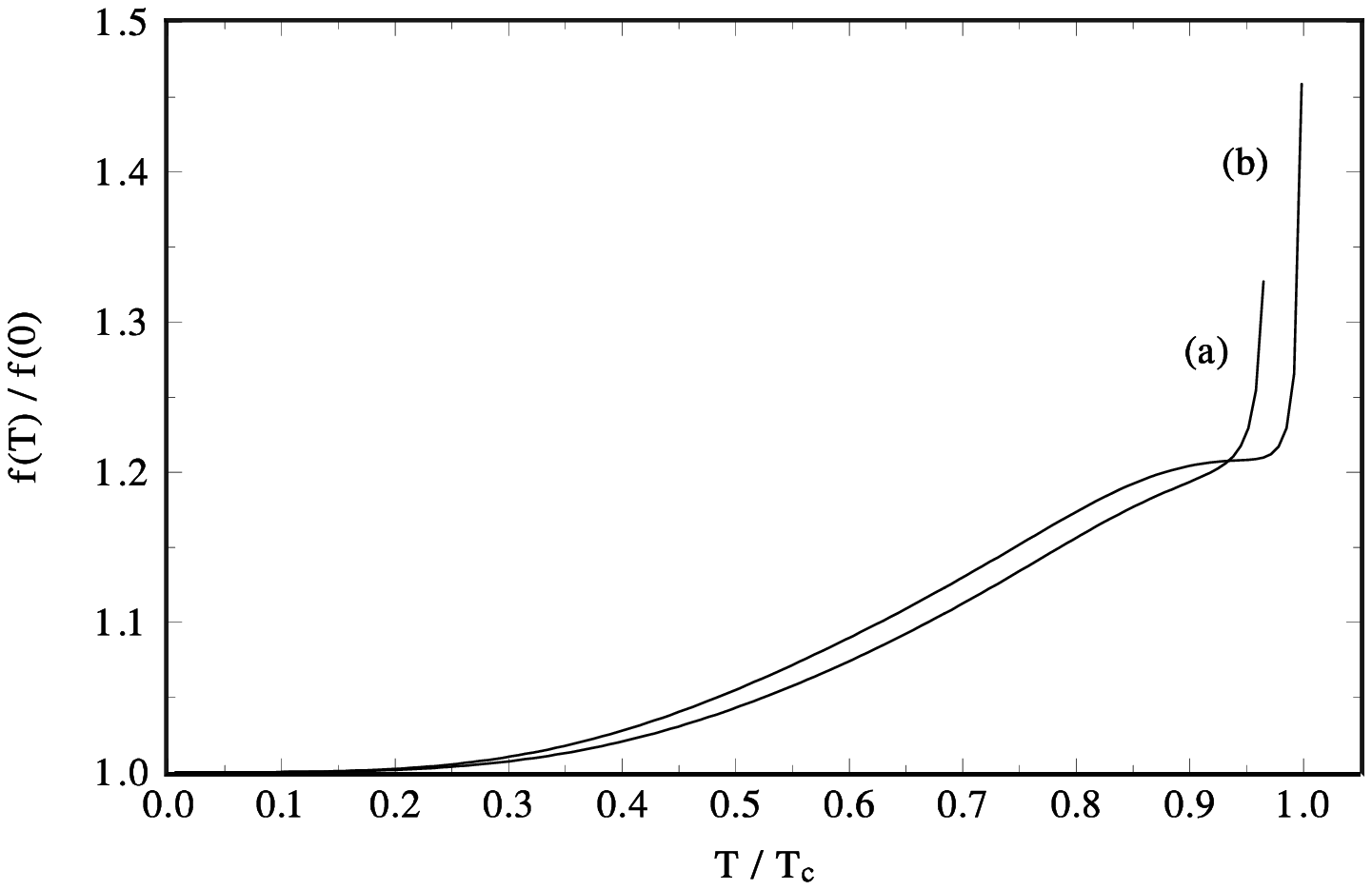}}
	}
	\caption{The ratio of the  couplings $f(T)/f(0)$ against $T/T_c$ in the vector channel ($\Upsilon$), curve (a), and in the pseudoscalar channel ($\eta_b$), curve (b).}
	\label{F}
\end{figure}

\section{RESULTS AND CONCLUSIONS}

We follow  closely the procedure employed previously to analyze charmonium in the vector channel \cite{JPSI}, as well as in the scalar and pseudoscalar channels \cite{ETAC}. The PQCD threshold $s_0(T)$ and the parameter $Q_0^2$ are obtained from the QCD ratio

\begin{equation}
	\frac{\varphi_N(Q_0^2,T)|_{QCD}}{\varphi_{N+1}(Q_0^2,T)|_{QCD}} = \frac{\varphi_{N+1}(Q_0^2,T)|_{QCD}}{\varphi_{N+2}(Q_0^2,T)|_{QCD}} \;,
\end{equation}

where $\varphi_N(Q_0^2,T)|_{QCD} = \varphi_N(Q_0^2,T)|_{PQCD} + \varphi_N(Q_0^2,T)|_{NP}$. A posteriori, this ratio and therefore $s_0(T)$, are fairly insensitive to $Q_0^2$ in a very wide range. These equations hold in the zero-width approximation, which remains valid even if the width were to increase with temperature by 3 orders of magnitude, say from $\Gamma(0) \simeq 100 \;{\mbox{KeV}}$ to $\Gamma(T) \simeq 300 \;{\mbox{MeV}}$. The hadron mass follows from

\begin{equation}
	\frac{\varphi_1(Q_0^2,T)|_{HAD}}{\varphi_2(Q_0^2,T)|_{HAD}} = \frac{\varphi_1(Q_0^2,T)|_{QCD}}{\varphi_2(Q_0^2,T)|_{QCD}} \,,
\end{equation}

and the width follows from

\begin{equation}
	\frac{\varphi_1(Q_0^2,T)|_{HAD}}{\varphi_3(Q_0^2,T)|_{HAD}} = \frac{\varphi_1(Q_0^2,T)|_{QCD}}{\varphi_3(Q_0^2,T)|_{QCD}} \,.
\end{equation}

Finally, the coupling is obtained e.g. from 

\begin{equation}
	\varphi_1(Q_0^2,T)|_{HAD} = \varphi_1(Q_0^2,T)|_{QCD}\;.
\end{equation}

We begin this procedure at $T=0$ in order to confront results with experimental data and thus check the accuracy of the method. As input values of the various parameters we use a  bottom-quark pole mass \cite{PDG}-\cite{mb} $m_b(m_b) = 4.65 \; {\mbox{GeV}}$, which gives a PQCD threshold $s_{th} = 4 m_b^2 = 86.5 \;{\mbox{GeV}}^2$, and a gluon condensate \cite{G2} 
$\left\langle\frac{\alpha_s}{\pi}G^2\right\rangle_{T=0} = 0.005 \; {\mbox{GeV}}^4$, allowing for a $50  \%$ uncertainty. In the vector channel ($\Upsilon$) the sum rules at $T=0$ give $s_0(0)= 95.3 \;{\mbox{GeV}}^2$ ($\sqrt{s_0(0)} = 9.76 \;{\mbox{GeV}}$), $M_V(0) = 9.6\;{\mbox{GeV}}$ to be compared with the experimental value $M_V(0)|_{EXP} = 9.46\;{\mbox{GeV}}$, $\Gamma_V(0) = 54  \;{\mbox{KeV}}$, identical to its experimental value, and $f_V(0) = 180 \;{\mbox{MeV}}$. These results are for $Q_0^2 = 10 \;{\mbox{GeV}}^2$; varying it in the range $Q_0^2 = 1 - 20 \;{\mbox{GeV}}^2$ changes the output by less than 5 \%. In the pseudoscalar channel ($\eta_b$) the solutions give $s_0(0)= 88.6 \;{\mbox{GeV}}^2$ ($\sqrt{s_0(0)} = 9.41 \;{\mbox{GeV}}$), $M_P(0) = 9.4\;{\mbox{GeV}}$ to be compared with the experimental value $M_P(0)|_{EXP} = 9.39\;{\mbox{GeV}}$, $\Gamma_P(0) = 50  \;{\mbox{KeV}}$, not known from experiment, and $f_P(0) = 90 \;{\mbox{MeV}}$. The parameter $Q_0^2$ was varied in the range $Q_0^2 = 0 - 10 \;{\mbox{GeV}}^2$, with stable results  as in the vector channel. One should notice that $s_0(0)$ in both channels is very close to the hadronic threshold $s_{th}|_{HAD} \simeq M_{V,P}^2$, as well as to the QCD threshold $s_{th}|_{QCD} = 4 \, m_b^2$. From experience in the charmonium system this means that at finite $T$ the PQCD threshold, $s_0(T)$, and  the hadron mass, $M(T)$, are expected to change very little with temperature. Once $s_0(T)$ decreases to $s_{th}$ close to $T=T_c$ the moment integrals vanish, thus the region above $T_c$ cannot be explored with this method.\\
Turning on the temperature, the only additional input quantity is the thermal gluon condensate for which we use a recent smooth fit \cite{recent} to the LQCD data of \cite{latticeG}

\begin{equation}
	\frac{\left\langle \alpha_s G^2\right\rangle_T}{\left\langle \alpha_s G^2\right\rangle_{T=0}} = 1 - a \left(\frac{T}{T_c}\right)^\alpha  \;,
\end{equation}

where $a= 1.015$, and $\alpha= 3.078$.  Results for the mass ratio, $M(T)/M(0)$ and for the PQCD threshold ratio, $s_0(T)/s_0(0)$ as a function of $T/T_c$ are shown in Figs. 1 and 2, respectively. As expected there is a very small change with increasing T, as $s_0(0)$ is quite close to threshold. Indeed, at $T/T_c \simeq 1$, $s_0(T)/s_0(0) \simeq 0.91 (0.98)$, for the vector (pseudoscalar) channel, translating in both cases into the same final value $s_0(T_c) = 4 m_b^2$. In contrast, the behaviour of the width and leptonic decay constant is quite different, as shown in Figs.3 - 5. We plot the ratio $\Gamma(T)/T$ versus $T/T_c$ for the $\Upsilon$ in Fig. 3, and for the $\eta_b$ in Fig.4, to facilitate a comparison with LQCD results which have used these axes \cite{BOTT}. There is a remarkable qualitative agreement between both methods, once account is taken of the different conceptual meanings and numerical values of the critical temperature. As in the charmonium channel \cite{JPSI}-\cite{ETAC}, these results, together with those for the leptonic decay constants shown in Fig. 5, strongly suggest the survival of $\Upsilon$ and $\eta_b$ beyond $T_c$.\\
\section{Acknowledgements}
This work has been supported in part by NRF (South Africa), Alexander von Humboldt Foundation (Germany), and FONDECYT (Chile) 11130056. 

\end{document}